# Grafting fluorescent nanodiamonds onto optical tips

A. Cuche,[a] A. Drezet,[a] J.-F. Roch,[b] F. Treussart,[b] S. Huant[a]

[a] Institut Néel, CNRS & Universié Joseph Fourier, BP 166, 38042 Grenoble, France
serge.huant@grenoble.cnrs.fr
[b] Laboratoire de Photonique Quantique & Moléculaire, Ecole Normale Supérieure de Cachan
& CNRS UMR 8537, Cachan, France

**Abstract.** We recently introduced an all-optical method for grafting onto the apex of an optical tip a single 20 nm nanodiamond with single color-center occupancy and used the resulting single-photon tip in scanning near-field imaging at room temperature, thereby achieving a genuine scanning single-photon microscopy working in ambient conditions. A variant of this method is described that allows for attaching several nanodiamonds onto the tip apex, releasing them all at once and finally recapturing them one by one by the scanning tip. This underlines the flexibility and powerfulness of our method and its variant that could be used in applications where a fixed number of selected optically active nano-objects requires positioning, or repositioning, at well defined locations with nanometer accuracy.

## 1 INTRODUCTION

Active optical tips for near-field optics are made of optically-active nano-objects ─ possibly only one ─ attached onto the apex of an optical fiber tip. Implementation of the latter in a near-field scanning optical (NSOM) environment simultaneously allows for positioning of the nano-object with nanometer accuracy and for its excitation by injecting an appropriate laser light into the optical fiber. Using solely the light emitted by the object for illumination purpose should pave the way to innovative studies in nano-optics at ultra-high spatial resolution, ultimately scaling down to the size of the emitting object. In the last decade, various nano-objects have been used for producing active tips such as single molecules [1], fluorescent diamond crystals in the 100 nm range [2], colloidal semiconductor quantum dots [3-5], rare-earth doped oxide nanoparticles [6], or integrated nano light-emitting devices [7].

Recently, we have introduced an all-optical method for grafting onto an optical tip a single nanodiamond with size in the sub-100 nm range, namely 20 nm, and with single color-center nitrogen-vacancy (NV) occupancy [8]. The single color-center acts as a single quantum emitter, i.e., a single-photon source, that is extremely photostable (no emission intermittency, no bleaching [9]) and works at room temperature. The use of the nanodiamond-based single-photon tip in NSOM achieves a scanning single-photon microscopy. The method used for attaching a diamond nanocrystal is simple and reliable. To better stress its versatility, we describe here a set of complementary experiments where several nanodiamonds are successively grafted one by one, released all at once, and then recaptured one by one by the scanning tip. Subsequent use of the functionalized tip in NSOM imaging shows that, despite this rather severe treatment, the tip has kept its imaging capability entirely.

## 2 GRAFTING A NANODIAMOND HOSTING A SINGLE NV-CENTER

Our method has been described in detail in Ref. 8. Briefly, an etched fiber-tip is coated with a positively charged polymer (poly-l-lysine), see Fig. 1(a), implemented in a tuning-fork [10] based NSOM where it is used in the excitation mode, and then scanned in the near-field of a cover slip onto which nanodiamonds (size < 50 nm) have been dispersed. The near-field fluorescence image allows for localizing fluorescent single nanodiamonds of a small size around 20 nm, as revealed by the topography image acquired simultaneously with the fluorescence [11]. During this localization step, the tip is kept at a safe altitude of 50 nm above the scanned surface so that no diamond is displaced or fortuitously captured by the tip. In addition, this allows for an accurate topography image acquisition that provides us with the crystal size. Once an isolated diamond nanocrystal has been selected, a new scan is launched and the scanning tip is pushed down at a lower altitude of 20 to 30 nm, and maintained there over a few scan lines surrounding the diamond location, in such a way that this very nanodiamond is captured by the tip: see Fig. 1(b). This efficient grafting occurs thanks to the negative charges located on the nanodiamonds' surface as a result of their chemical treatment at the stage of their dispersion in colloidal suspensions. The diamond attachment is confirmed by subsequent spectroscopic measurements [12] and photon-intensity correlation measurements [13] allows the quantification of the number of NV centers embedded in the nanoparticle [8,9,14].

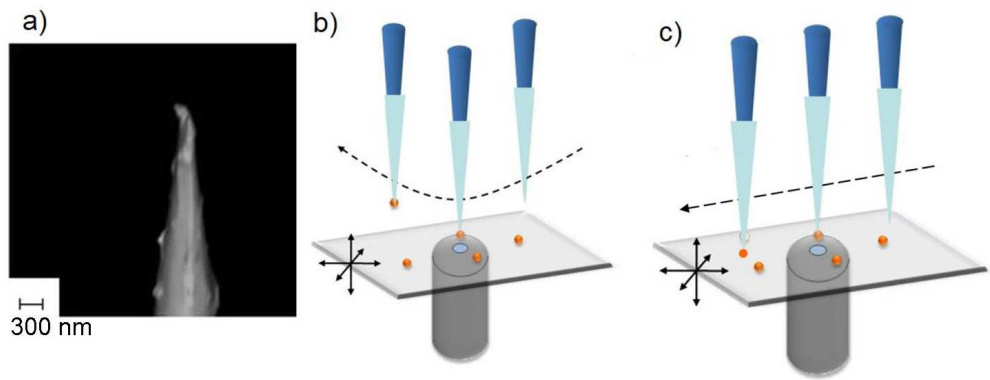

Fig. 1. a) A scanning electron micrograph of an optical tip coated with poly-l-lysine confirming that the polymer covers the tip entirely, including its *apex*. b) Scheme of the process used for grafting a selected single nanodiamond. The dashed arrow illustrates the tip trajectory: the tip is approached to the surface from a cruise altitude of about 50 nm down to an altitude of about 20 to 30 nm at the very moment when it is facing the selected nanocrystal. c) Simplified variant of the former process. Here, the tip continuously scans at a low height of about 20 nm in such a way that every encountered nanocrystal is embarked.

## 3 FISHING, FREEING AND FISHING BACK SEVERAL UNSELECTED DIAMONDS

The process summarized above is all-optical in the sense that it is achieved entirely in a single NSOM environment. There is no need to employ additional manipulators or electron microscopes to control the grafting [15,16]. The optical tip, for instance, is successively used to locate and identify the very nanodiamond to be grafted, to graft it, and in the subsequent optical imaging. In contrast with bare optical tips, i.e., not coated with poly-l-lysine [17], having seized nanodiamonds during scanning, the graft is here strong enough to use the resulting active tip as a nanosource of light in a near-field microscopy experiment [8].

A simpler variant of this process consists in scanning the trapping tip directly at a small altitude of around 20 nm above the surface as shown schematically in Fig. 1(c). This way, a larger number of unselected nanodiamonds, including possible non fluorescent objects, are captured. Here, it is not possible to discern the individual objects on the topography image since they are taken away prior to their topographic characterization. Each trapping event of a fluorescent nanodiamond manifests itself in the concomitant embarking of its fluorescence as seen both in the image in Fig. 2(a) and in the related cross-cut in Fig. 2(b). In particular, single trapping events (labeled "+1" in Fig. 2(b)) produce step-like increases in the fluorescence intensity of the functionalized tip of approximately 5000 counts/pixel (5 kcount/px). This amounts to the smallest intensity level detected for the particular nanodiamond sample used in this work under the experimental conditions of Fig. 2. Although a single NV-center occupancy can only be asserted from photon-intensity correlation measurements, we take this increase of 5 kcount/px as a hint that a single color-center has indeed been captured by the tip. Interestingly enough, a signal increase twice as intense is observed along the central horizontal line in Fig. 2(a) as confirmed in the cross-section in Fig. 2(b) (peak labeled "+2 ?"). This can either be due to a trapping of a diamond with double NV-occupancy or to the trapping of two objects with single-NV occupancy. Our data are unable to discriminate between both *scenarios*. Finally, we note in Fig. 2 that, within the experimental accuracy, the intensity emitted by the functionalized tip remains equal to 20 kcount/px for the rest of the image. This roughly corresponds to the emission level of four NV centers. These centers are neutral, as confirmed by the emission spectrum of the tip recorded after scanning (not shown, see Ref. 8 for a typical spectrum of a diamond-based optical tip).

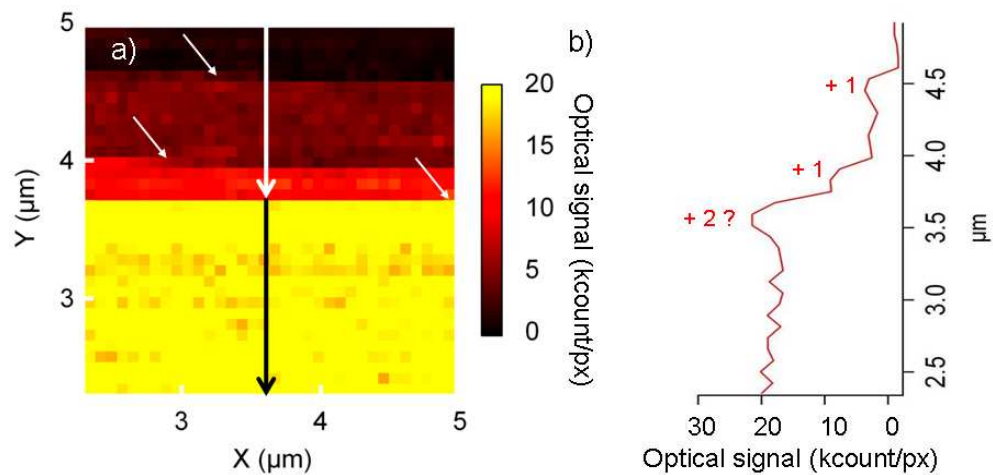

Fig. 2. a) Fluorescence image acquired in the spectral range of the NV emission showing the successive trapping of several nanodiamonds during the tip scan close ($\approx$ 20 nm) to the surface. Short arrows tentatively point towards the trapping locations. The tip is scanned from right to left (fast scan direction) and from top to bottom (slow scan). Each trapping event translates into a persistent increase in the tip-nanodiamond assembly fluorescence. Experimental parameters are: 2.5($\mu$m)$^2$ scanning area, excitation at 488 nm, 2.5 µW excitation power at the tip apex, 70 ms acquisition time per pixel. b) Intensity cross-section along the central vertical arrow (parallel to the slow-scan direction) showing three successive trapping events.

The graft is strong enough to use the functionalized tip for NSOM imaging provided that the probe tip is kept at an altitude of 20 to 30 nm above the surface. However, touching the surface with the tip, even in a gentle manner, results in the release of the captured nanodiamonds due to a contact force in excess of a few tens of nN [10]. This immediately translates into a complete loss of fluorescence of the tip-nanodiamond assembly. We could next check that scanning again over the contact zone where the diamonds have been released allows for a recapture of every nanodiamond one by one along a process similar to that shown in Fig. 2.

## 3 NEAR-FIELD IMAGING WITH A RE-FUNCTIONALIZED TIP.

The interesting point to stress is that a "re-functionalized" tip, i.e. a tip that has first released all its captured nanodiamonds and then re-captured some of them is useable in NSOM imaging as it was prior to the nanodiamond release [8]. A demonstration of this ability is given in Fig. 3(a) which depicts a transmission NSOM image of a test sample (made of gold discs of diameters 400 nm and thicknesses 40 nm patterned on a fused silica cover slip) acquired with the NV-fluorescence light emitted by a tip that has recaptured exactly two nanodiamonds.

As can be seen in Fig. 3(a), the contrast is sufficient not only to reveal the gold pattern as dark non-transmitting spots, but also some smaller scale defects that are visible in the electron micrograph of Fig. 3(b) (such as for example a tiny hole at the center of the upper right gold disc). From Fig. 3(a), it can be estimated that the spatial resolution is in the 150 nm range, i.e., much better than with the initial uncoated tip which offers diffraction-limited resolutions in the 400 nm range [11]. More details about the spatial resolution potentially offered by nanodiamond-based tips can be found in Ref. 8.

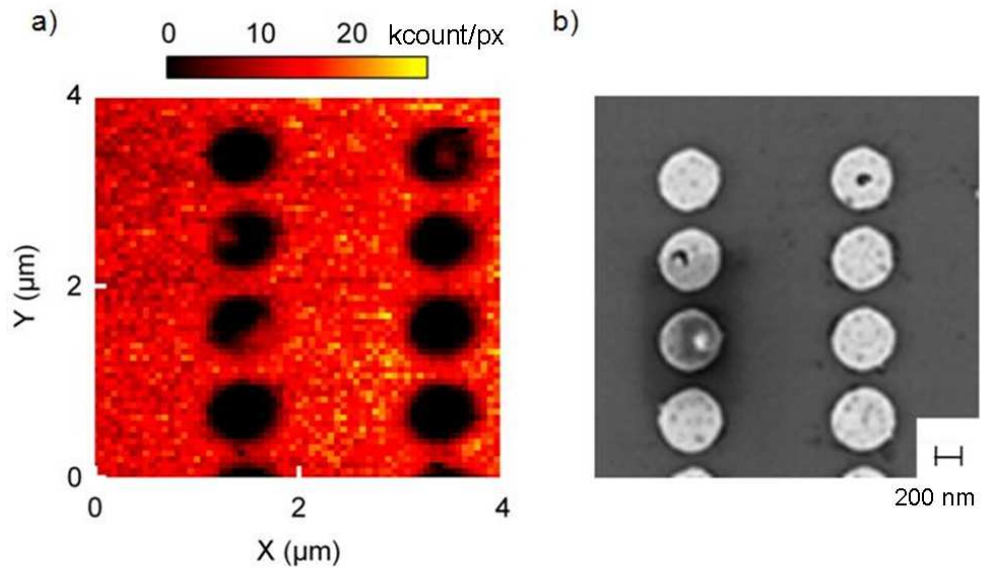

Figure 3: a) Transmission NSOM image of gold discs patterned on a cover slip acquired with the fluorescence emitted by a "re-functionalized" tip as source of light. Here, the detected light has been restricted by spectral filtering to the emission band of the neutral NV center. Experimental parameters are: $2(\mu m)^2$ scanning area, excitation at 488 nm, 60 μW excitation power at the tip apex, 60 ms acquisition time per pixel. b) Scanning electron micrograph of the same area.

## 4 SOME PROSPECTS

To summarize, we have described a simple optical method to capture several fluorescent nanodiamonds with an optical tip, to release all of them, and to possibly recapture some of them. Combined with the controlled method described in Ref. 8 which allows attaching one, and only one, well-selected nanodiamond, we believe our work might be useful in various studies such as accurate launching of surface plasmons [18,19], manipulation and positioning of a quantum object with nanometer accuracy in the vicinity of mechanical resonators or plasmonic devices for quantum optomechanics [20] and plasmonics [21], or high-resolution magnetometry [22], all of this at room temperature.

## Acknowledgments

We are grateful to J.-F. Motte for the optical tip manufacturing, to J.-P. Boudou and T. Sauvage for the fluorescent nanodiamond preparation, and to Y. Sonnefraud and O. Faklaris for their contribution to early stages of this work. The PhD grant of AC by the Région Rhône-Alpes through "Cluster MicroNano" is gratefully acknowledged. This work was supported by the European Commission through the EQUIND and NEDQIT projects and by Agence Nationale de la Recherche, France, through the NADIA, PROSPIQ, and NAPHO projects.